\documentclass[prb,preprint,amsmath,amssymb,showpacs,superscriptaddress,floatfix,longbibliography]{revtex4-1}
\usepackage[breaklinks=true,colorlinks,citecolor=blue,linkcolor=blue,urlcolor=blue]{hyperref}

\usepackage{epsfig,mathrsfs,color,latexsym,subfigure,marginnote,gensymb,}
\usepackage{graphicx}
\usepackage{makecell}

\renewcommand{\BibitemShut}[1]{}

\begin{document}

\title{Intrinsic magnetism in monolayer transition metal trihalides: a comparative study}% Force line breaks with \\
\author{Shalini Tomar}
\affiliation{Department of Physics, Bundelkhand University, Jhansi 284128, India}
\affiliation{Department of Electrical Engineering, IIT Kanpur, Kanpur 208016, India}
\author{Barun Ghosh}
\affiliation{Department of Physics, Indian Institute of Technology , Kanpur 208016, India}
\author{Sougata Mardanya}
\affiliation{Department of Physics, Indian Institute of Technology , Kanpur 208016, India}
\author{Priyank Rastogi}
\affiliation{Department of Electrical Engineering, IIT Kanpur, Kanpur 208016, India}
\author{B. S. Bhadoria}
\affiliation{Department of Physics, Bundelkhand University, Jhansi 284128, India}
\author{Yogesh Singh Chauhan}
\affiliation{Department of Electrical Engineering, IIT Kanpur, Kanpur 208016, India}
\author{Amit Agarwal}
\email{amitag@iitk.ac.in}
\affiliation{Department of Physics, Indian Institute of Technology , Kanpur 208016, India}
\author{Somnath Bhowmick}
\email{bsomnath@iitk.ac.in}
\affiliation{Department of Materials Science and Engineering, IIT Kanpur, Kanpur 208016, India}

\date{\today}% It is always \today, today,

\begin{abstract}
Two dimensional magnetic materials, with tunable electronic properties could lead to new spintronic, magnetic and magneto-optic applications. Here, we explore intrinsic magnetic ordering in two dimensional monolayers of transition metal tri-halides (MX$_3$, M = V, Cr, Mn, Fe and Ni, and X = F, Cl, Br and I), using density functional theory. We find that other than FeX$_3$ family which has an anti-ferromagnetic ground state, rest of the trihalides are ferromagnetic. Amongst these the VX$_3$ and NiX$_3$ family are found to have the highest magnetic transition temperature, beyond the room temperature.  In terms of electronic properties, the tri-halides of  Mn and Ni are either half metals or Dirac half metals, while the tri-halides of V, Fe and Cr are insulators. Among all the trihalides studied in this paper, we find the existence of very clean spin polarized Dirac half metallic state in MnF$_3$, MnCl$_3$, MnBr$_3$, NiF$_3$ and NiCl$_3$. These spin polarized Dirac half metals will be immensely useful for spin-current generation and other spintronic applications.
\end{abstract}

\maketitle

%\tableofcontents

\section{\label{sec:level1}Introduction}

Recent discovery of ferromagnetism in single layer CrI$_3$ and Cr$_2$Ge$_2$Te$_6$, has opened up the field of intrinsic magnetism in 2D materials to intense exploration.\cite{Nature_Cri3,Nature2} Magnetism in ultra-thin two dimensional (2D) materials opens up the exciting possibility of atomically thin and low dimensional magnetic storage, sensing, spintronic and optical devices. \cite{Huang2018,Jiang2018,Burch2018,2053-1583-4-2-024009,Seyler2018,YARMOHAMMADI2018103,IGOSHEV20123601,Mao_2018,C1CP22719J,SUN201746} Additionally, they can also be used to break time-reversal symmetry in 2D stacks for valley-tronics.\cite{Jiang2018} Indeed, electrostatic gate control of magnetism in both CrI$_3$ and Cr$_2$Ge$_2$Te$_6$ has already been demonstrated,\cite{Huang2018,Jiang2018,Burch2018,2053-1583-4-2-024009} along with spin-filter device with giant tunnelling magnetoresistance.\cite{Song2018,Klein2018,acs.nanolett.8b01552,CRI3_TMR} In addition to their applicability, 2D magnetism also opens the door to new quantum state of matter, existence of which was prohibited by the Mermin Wagner theorem.\cite{Mermin_1966_PRL}  

To understand the prohibition of spin-ordering in 2D systems by the Mermin Wagner theorem,\cite{Mermin_1966_PRL}  let us have a look at the simplest model describing spin-spin interaction in a 2D system,  
\begin{equation} \label{eq1}
H =  - J_{\parallel} \sum_{kl} (S^x_k S^x_{k+l} + S^y_k S^y_{k+l}) - J_{\perp} \sum_{kl} S^z_k S^z_{k+l}~.
\end{equation}
Here $S^m_k$ denotes the $m = x,y,z$ component of the spin at site $k$, and the sum over $k,l$ runs over the nearest neighbors. In Eq.~\eqref{eq1}, $J_{\parallel}$ and $J_{\perp}$ are the in-plane ($x-y$ plane) and out of plane spin exchange energy, respectively. If $J_{\perp} = 0$ and $J_{\parallel} \neq 0$, the Hamiltonian reduces to the X-Y Ising-spin model, and for $J_{\perp} = J_{\parallel} \neq 0$ the model reduces to the isotropic Heisenberg spin model. The Mermin Wagner theorem\cite{Mermin_1966_PRL} forbids finite temperature magnetic ordering in the isotropic X-Y, as well as the isotropic Heisenberg spin model  (both classical and quantum) in one and two dimensions. This is a consequence of quantum and thermal fluctuations dominating over the spin-ordering phenomena. However, spontaneous magnetic ordering can still emerge in SU(2) symmetry broken cases such as the anisotropic Heisenberg model ($J_{\parallel} \neq J_{\perp}$), or the Ising model. In fact, Onsager demonstrated the theoretical possibility of a finite temperature magnetic transition in the two dimensional Ising model as early as 1944.\cite{Onsager1944} The spin ordering observed using high-sensitivity microscopy (polar magneto-optical Kerr effect microscopy) technique in mono-layered CrI$_3$ is indeed found to be Ising like (below 45K).\cite{Nature_Cri3,Samarth2017,CrI31} In case of Cr$_2$Ge$_2$Te$_6$, while no magnetic ordering is observed in the monolayer (at least till 4.7 K), the magnetic ordering in the bi-layer is described by the Heisenberg model with additional magnetic anisotropies \cite{Nature2,Samarth2017} (a term $\propto \sum_{k} (S^z_k)^2$). 

\begin{figure*}
\includegraphics[width=\linewidth]{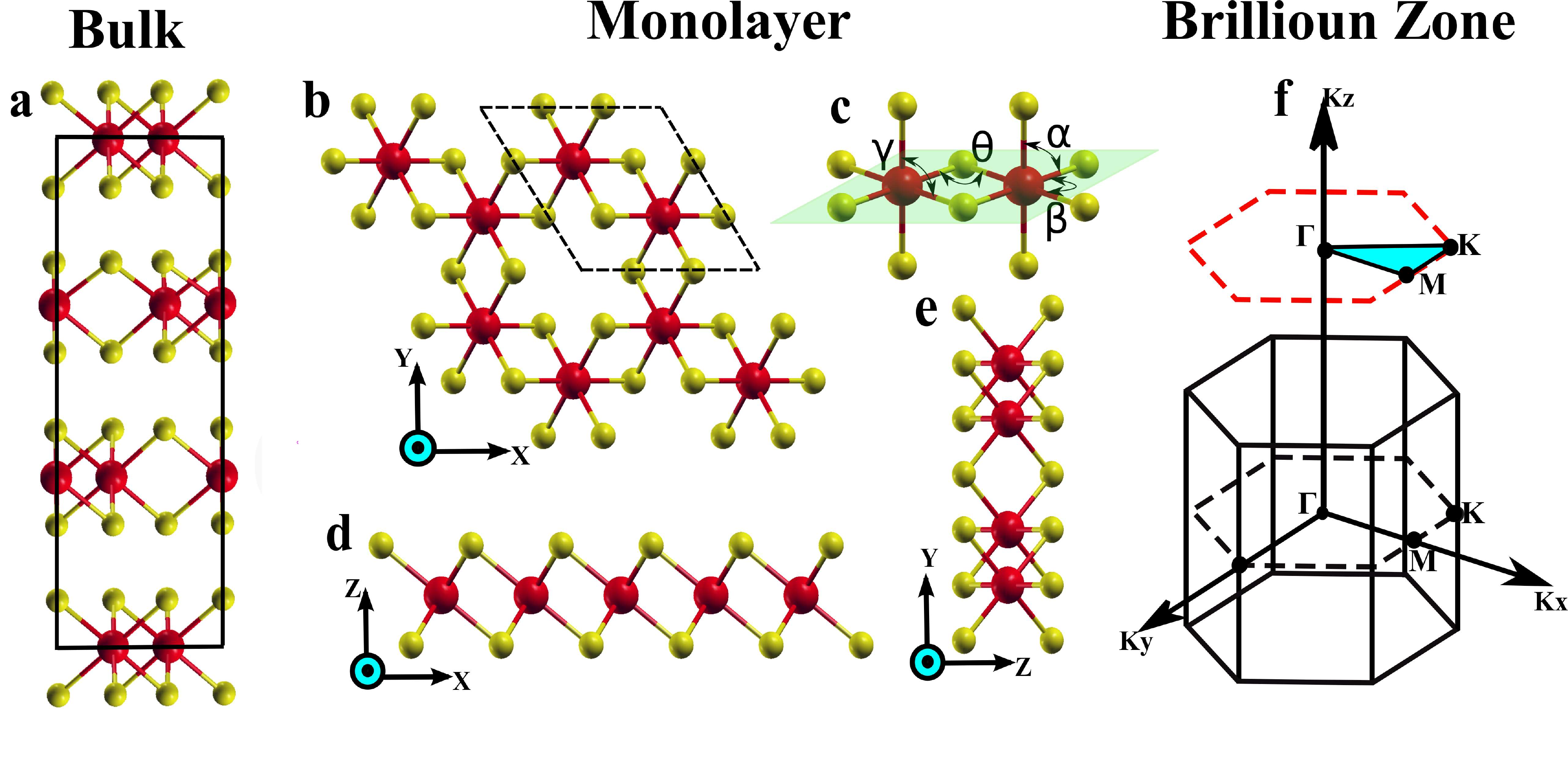}
\caption{Crystal structure and Brillouin zone of bulk and monolayer MX$_3$. (a) BiI$_3$ type rhombohedral unit cell of bulk MX$_3$. (b) Top view of monolayer MX$_3$, with the hexagonal unit cell marked by the dashed line. (c) Six X atoms (yellow) forming an octahedron surrounding the M atom (red). (d) and (e) show different side views of MX$_3$ monolayers. (f) Brillouin zones of bulk (solid line) and monolayer (dashed line) MX$_3$, corresponding to the unit cells shown in (a) and (b).} 
\label{Fig1}
\end{figure*}

Following CrI$_3$ and Cr$_2$Ge$_2$Te$_6$, 2D magnetism has also been experimentally reported in FePS$_3$\cite{2053-1583-3-3-031009,Lee2016} (Ising-type anti-ferromagnet) and MnSe$_2$\cite{OHara2018} (ferromagnet), among others. In a more significant development, room temperature magnetism has been recently demonstrated in VSe$_2$ monolayers on van der Waals substrates.\cite{Bonilla2018} Contrary to CrI$_3$ and Cr$_2$Ge$_2$Te$_6$, VSe$_2$ is paramagnetic in the bulk form. In addition to these, there have been theoretical predictions of magnetism in several other materials based on density functional theory and Monte-Carlo simulations.\cite{PhysRevMaterials.2.081001}  These include several monolayers of easily exfoliable metal di-halides MX$_2$ (M = Fe, Co, Ni, Mn, V, and X=B, Cl, I),\cite{MX2_1} and layered ternary compounds CrSiTe$_3$, CrGeTe$_3$ and Cr$_2$Ge$_2$Te$_6$ of the telluride family.\cite{CrXTe3,5024576}  Several of the transition metal tri-halides such as CrX$_3$,\cite{CrI32,C5TC02840J} OsCl$_3$,\cite{OsCl3_v1} VCl$_3$,\cite{Zhou_2016,VCl3} NiCl$_3$,\cite{NiCl3} RuX$_3$, \cite{RuCl3_PCCP, RuCl3_JMCC, RuCl3_Nanoletter,ersan2019} and MnX$_3$ (X = F, Cl, Br,I)\cite{MnCl3} are also expected to have either a ferromagnetic or an anti-ferromagnetic ground state described by an Ising spin model. 

Motivated by this recent interest in 2D magnetic materials, we present a comparative study of magnetism in different transition metal tri-halide monolayers. More specifically, we investigate the crystal structure, electronic and magnetic properties for 20 different MX$_3$ compounds (M$=$ V, Cr, Mn, Fe and Ni, while X$ =$ F, Cl, Br, or I), using density functional theory. Not only we find several ferromagnetic  2D materials (V, Cr, Mn and Ni tri-halides), but also anti-ferromagnetic monolayers (Fe tri-halides), not reported so far in the literature. We also identify several spin polarized Dirac half-metals among the family of transition metal tri-halide monolayers. These have a large band-gap in one spin channel and a Dirac cone in the other, with Fermi velocities comparable to graphene. Such materials are ideal for making spintronic and other nano-electronic devices. 

The paper is organized as follows: computational details are described in Sec.~\ref{cd}. Results are presented along with detailed discussions in Sec.~\ref{rd}. It contains a description of the crystal structure, analysis of binding energy in the context of stability of the monolayers and discussions on electronic and magnetic properties of transition metal tri-halide monolayers. Our findings are summarized in Sec.~\ref{sec:level6}.

\section{\label{cd}Computational Method}
Using density functional theory (DFT), we calculate the structural, electronic and magnetic properties of MX$_3$ monolayers.  We use a plane-wave basis set with the cutoff energy of 80 Ry, and projector-augmented wave pseudo potentials, as implemented in the Quantum Espresso (QE) and VASP package.\cite{QE,paw,PhysRevB.59.1758,PhysRevB.54.11169} The electron-electron exchange correlation is treated within the framework of the generalized gradient approximation (GGA), as proposed by Perdrew-Burke-Ernzerhof (PBE).\cite{PBE} Brillouin zone integrations are performed using a $k$-point grid of $32\times32\times1$. Perpendicular to the monolayers, a vacuum of 15 \AA~ is applied to prevent the spurious interactions among replica images. All the structures are fully relaxed until the forces on each atom are less than 10$^{-3}$ Ry/au and the energy difference between two successive ionic relaxation steps is less than 10$^{-4}$ Ry. Note that, electronic band structure calculations based on GGA-PBE approximation yield consistent results in case of Quantum Espresso and VASP, while only the latter is used to perform more accurate hybrid functional (HSE06) based calculations.\cite{doi:10.1063/1.1564060}

\section{\label{rd} Result and Discussion}
\subsection{Crystal Structure}
Bulk transition metal tri-halides (MX$_3$) are known to exist in the form of layered materials bonded by weak van der Waals interactions. These materials are generally found either in monoclinic AlCl$_3$ or rhombohedral BiI$_3$ crystal structure.\cite{crystal_struct} A bulk unit cell of BiI$_3$ type is shown in Fig.~\ref{Fig1}(a), where monolayers are stacked in ABC sequence.  Several of transition metal tri-halides are predicted to have low values of the cleavage energy, based on  first principle calculations.\cite{CrI31,C5TC02840J,CrI32,0295-5075-114-4-47001,Wang_JPCM_2011,Lado_2D_2017} Thus, in principle they can be exfoliated as atomically thin 2D materials from their bulk phases. 

Recently, a single magnetic layer of CrI$_3$ has been successfully exfoliated from its bulk phase,\cite{Nature_Cri3} encouraging the exploration of monolayers of the tri-halide family of magnetic materials in 2D. In case of individual monolayers, the smallest repeat segment is a hexagonal unit cell, having two formula units of MX$_3$ in it [see Fig.~\ref{Fig1}(b)]. Each M atom is coordinated to six X atoms in an octahedral coordination, as shown in Fig.~\ref{Fig1}(c). Monolayers are not atomically flat and a layer of metal atoms are sandwiched between two layers of halide atoms, as shown in Fig.~\ref{Fig1}(d) and (e). The Brillouin zones (BZ) of the  bulk (solid line) and monolayer (dashed line) MX$_3$ is illustrated in Fig.~\ref{Fig1}(f). The {\it ab-initio} optimized values of the lattice constants, bond lengths (M-X) and bond angles (M-X-M) of 20 different MX$_3$ structures (M $=$ V, Cr, Mn, Fe and Ni, while X$=$ F, Cl, Br, or I) are reported in Table~\ref{Table1}. We find that the fluorides have slightly higher value of $\theta$ (lying between $\sim 99^\circ-104^\circ$) than that of chlorides, bromides and iodides (lying between $\sim 92^\circ-94.5^\circ$). As expected, bond lengths and lattice constants depend on the size of the constituent atoms and increase with the size of the halogen atoms and our numbers are consistent with the values reported in the literature.\cite{NiCl3,MnCl3,C8NR03322F,VCl3}

\begin{figure}
\centering
\includegraphics[width=\linewidth]{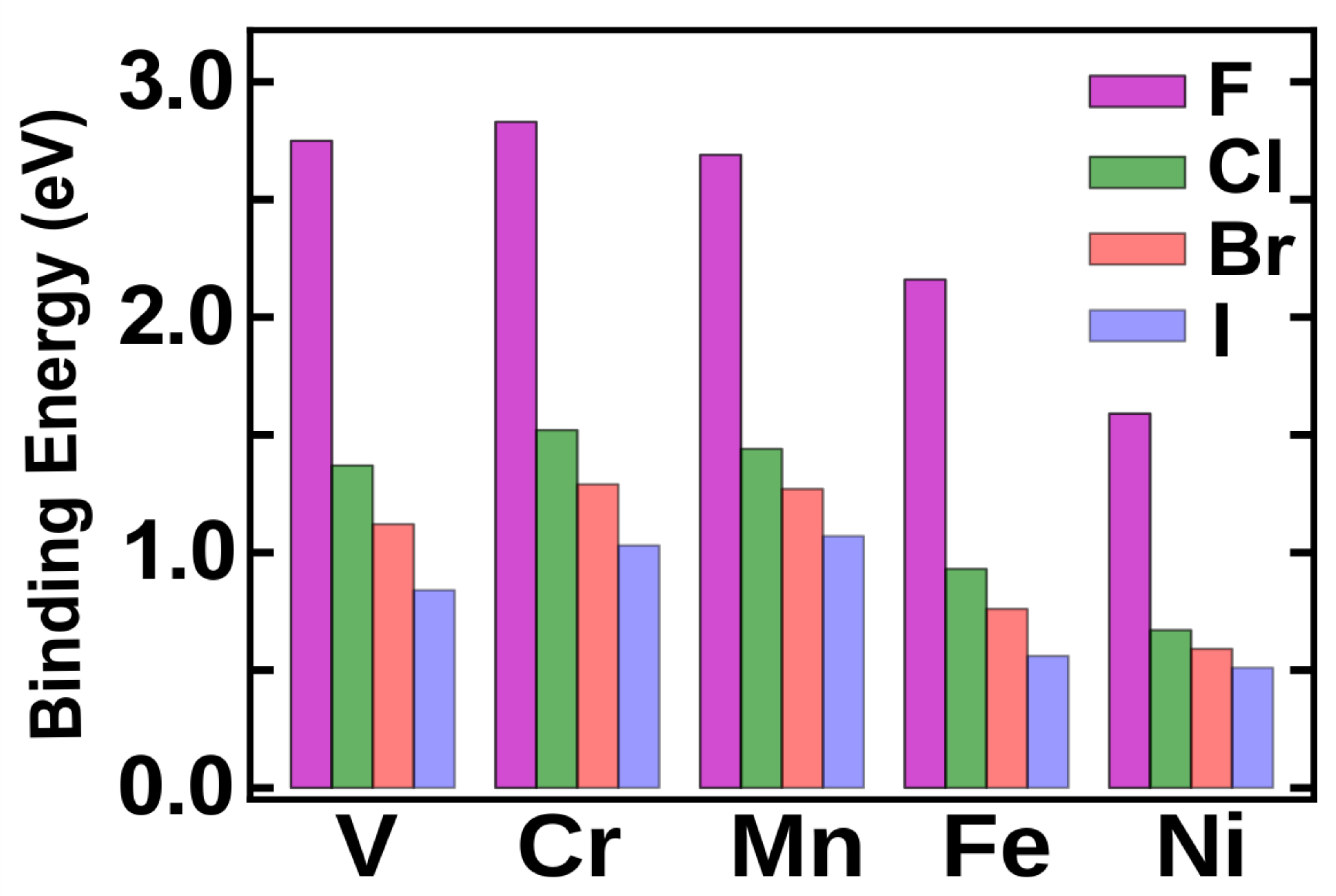}
\caption{ A comparison of calculated binding energies of MX$_3$ monolayers. Irrespective of the transition metals, fluorides have the highest binding energy among the halides.} 
\label{Fig2}
\end{figure}

%%&&&&&&&&&&&&&&&&&&&&&&&&&&&&&&&&&&&&&&&&&&&&&&&&&&&&&&&&&&&&&&&&&&&&&&&
\begin{table}
\renewcommand{\arraystretch}{0.7} % this reduces the vertical spacing between rows
\linespread{0.7}\selectfont
\small
\centering
\caption{Calculated lattice constant (a = b), M-X bond length (d$_{M-X}$), bond angles, and binding energy (BE) of MX${_3}$ monolayers. Different bond angles ($\alpha$, $\beta$, $\gamma$ and $\theta$) are defined in Fig.~\ref{Fig1}(c).}
\begin{tabular}%{c c c c c }
{p{1.0cm}p{1.0cm}p{1.0cm}p{3.8cm}p{1.0cm}}
\hline
\hline
\vspace{0.1cm}\\
\leftline{MX$_3$} & a(\AA) & d$_{M-X}$(\AA)&  \centering{Bond-angle}  ~~~$\alpha$~~~~~~$\beta$~~~~~~$\gamma$~~~~~~$\theta$ & BE (eV) \vspace{0.1cm}\\ 
\hline \\
%\vspace{0.2 cm} \\
%
%\hline 
\leftline{VF$_{3}$}  & 5.34   & 1.95  &~95.5~~75.8~~95.0~~104.0   & 2.756 \\    
\leftline{VCl$_{3}$} & 5.97   & 2.39  &~93.1~~87.9~~89.5~~92.0     & 1.370 \\  
\leftline{VBr$_{3}$} & 6.33   & 2.54  &~92.7~~87.8~~89.7~~92.1      & 1.127 \\ 
\leftline{VI$_{3}$}  & 6.86   & 2.75  &~91.9~~88.0~~90.0~~91.9     & 0.840 \\
\hline \\
%\vspace{0.1 cm} \\
\leftline{CrF$_{3}$}  & 5.14   & 1.92 &~90.4~~79.1~~95.4~~100 &2.838  \\ 
\leftline{CrCl$_{3}$} & 5.97   & 2.35 &~93.1~~85.8~~90.6~~94.0 &1.520   \\ 
\leftline{CrBr$_{3}$} & 6.34   & 2.51 &~92.3~~86.6~~90.5~~93.3 &1.299    \\ 
\leftline{CrI$_{3}$}  & 6.85   & 2.71 &~91.6~~86.5~~90.9~~93.0 &1.039   \\
\hline \\
%\vspace{0.1 cm} \\
\leftline{MnF$_{3}$}  & 5.37   & 1.99   &~95.5~~77.6~~93.8~~102.0      &2.697   \\ 
\leftline{MnCl$_{3}$} & 6.10   & 2.40   &~94.9~~85.8~~89.8~~94.0       &1.445 \\ 
\leftline{MnBr$_{3}$} & 6.39   & 2.55   &~93.1~~87.4~~89.7~~92.5       &1.274 \\ 
\leftline{MnI$_{3}$}  & 6.85   & 2.77   &~92.0~~88.8~~89.5~~91.0       &1.070 \\ 
\hline \\
%\vspace{0.1 cm} \\
\leftline{FeF$_{3}$}   & 5.21   &1.97   &~93.1~~80.8~~93.3~~99.1      &2.160   \\ 
\leftline{FeCl$_{3}$}  & 6.02   &2.40   &~93.3~~87.4~~89.7~~92.5       &0.931  \\ 
\leftline{FeBr$_{3}$}  & 6.42   &2.57   &~92.9~~87.8~~89.6~~92.1       &0.767   \\ 
\leftline{FeI$_{3}$ }  & 6.97   &2.78   &~92.0~~87.4~~90.2~~92.5     &0.561  \\ 
\hline \\
%\leftline{CoF$_{3}$}             &5.201   & 1.9507      & 100.6  & M      &   \\ 
%\leftline{CoCl$_{3}$}            &5.794   &2.2580       & 95.6   & 1.165  &  \\ 
%\leftline{CoBr$_{3}$}            &6.127   &2.4079       & 94.5   & 1.033 & \\ 
%\leftline{CoI$_{3}$}             &6.69    &2.6014       & 95.8   & 0.886 & \\ 
%\hline 
\leftline{NiF$_{3}$}  &5.00   &1.87        &~90.7~~79.3~~95.2~~100.5      & 1.596   \\                                 
\leftline{NiCl$_{3}$} &5.82   &2.29        &~93.0~~85.8~~90.6~~94.5       & 0.679  \\ 
\leftline{NiBr$_{3}$} &6.16   &2.44        &~92.7~~86.8~~90.3~~93.0       & 0.595   \\ 
\leftline{NiI$_{3}$}  &6.64   &2.66        &~92.3~~87.3~~90.1~~91.8       & 0.519  \\             	 
\hline
\hline
\end{tabular}
\label{Table1} 
\end{table}

\subsection{Binding Energy} 
Phonon dispersion calculations have already established the stability of several of the MX$_3$ monolayers. \cite{VCl3,MnCl3,C5TC02840J,NiCl3}
%The dynamical stability of several MX$_3$ monolayers have already been established  based on phonon dispersion calculations.\cite{VCl3,MnCl3,C5TC02840J,NiCl3} 
This strengthens the possibility of the existence of such a family of atomically thin and possibly magnetic 2D materials in free standing form. Some idea regarding the relative stability of the monolayer metal tri-halides can be obtained by comparing their binding energies, defined as,\cite{C7TC03003G,BI_3}
\begin{equation}
E_b=\frac{2E_M + 6E_X - E_{MX_3}}{8},
\end{equation}
where E$_M$, E$_X$, and E$_{MX{_3}}$ are the total energies of the transition metal atom (M), halogen atom (X), and transition metal tri-halide monolayer (MX$_3$). According to our calculations, $E_b$ of all the monolayers listed in Table~\ref{Table1}, are positive and lie in the range from 0.519 to 2.838 eV per atom.   Note that, the binding energy for the cobalt tri-halide family are found to be negative (not shown in the table), indicating inherent instability of those structures and thus, they are not considered in this article. The calculated values of $E_b$ listed in Table~\ref{Table1} are also shown in a bar graph for ease of comparison in Fig.~\ref{Fig2}. Evidently, the fluoride MX$_3$'s have the highest binding energies compared to other halides. We find that the $E_b$ decreases with increasing size of the halogen atom. Also note that  V, Cr and Mn tri-halides are likely to be more stable as compared to the Fe and Ni tri-halides.

Thus, our structural analysis predicts the exciting possibility of the existence of several magnetic monolayers in the family of transition metal tri-halides, all having similar crystal structure. In fact the experimental discovery of 2D magnetic CrI$_3$ monolayers might just be the beginning of an exciting journey with magnetism in 2D transition metal tri-halides and other materials. Some of the transition metal tri-halides are already predicted to have magnetic ground state \cite{NiCl3,MnCl3,C8NR03322F,VCl3}. In the next section, we do an exhaustive study of electronic and magnetic properties of all the 20 MX$_3$ monolayers listed in Table~\ref{Table1}. 

%\begin{figure}
%\centering
%\includegraphics[width=\linewidth]{fig3.pdf}
%\caption{The energy difference between nonmagnetic (NM) and ferromagnetic (FM) state ($E_{NM}-E_{FM}$, pink blocks) and nonmagnetic and antiferromagnetic state ($E_{NM}-E_{AFM}$, purple blocks) for MX$_3$ monolayers. A positive value indicates the stability of the magnetic ground state relative to the NM ground state.} 
%\label{Fig3}
%\end{figure}

\begin{table}
\renewcommand{\arraystretch}{0.7} % this reduces the vertical spacing between rows
\linespread{0.7}\selectfont
\small
\centering
\caption{Calculated value of exchange coupling constant (J), absolute magnetic moment per M atom (m$_A$), total magnetic moment per unit cell (m$_T$)  and bandgap (E$_g$) for MX${_3}$ monolayers.}
%\begin{tabular}{c c c c c c c c } 
\begin{tabular}{p{1.5cm}p{1.5cm}p{1.5cm}p{1.5cm}p{1.5cm}}
\hline
\hline \\
\leftline{MX$_3$}  & J (meV) & m$_A$ ($\mu{_B}$) & m$_T$ ($\mu{_B}$) & E$_g$ (eV) \\ 
%\vspace{0.1cm}
\hline \\
\leftline{VF$_{3}$}   & 32.25    & 1.89 & 4 & 3.23 \\    
\leftline{VCl$_{3}$}  & 21.60    & 1.96 & 4 & 2.51 \\  
\leftline{VBr$_{3}$}  & 18.43    & 2.03 & 4 & 2.12 \\ 
\leftline{VI$_{3}$}   &17.78     & 2.17 & 4 & 1.26 \\
\hline \\
\leftline{CrF$_{3}$}  &0.96      &2.86&6&5.09    \\ 
\leftline{CrCl$_{3}$} &0.85     &2.93&6&3.84 \\ 
\leftline{CrBr$_{3}$} &2.57     &3.03&6&2.87    \\ 
\leftline{CrI$_{3}$}  &1.70      &3.15&6&1.85 \\
\hline \\
\leftline{MnF$_{3}$}  &8.47     &3.87&8 &0 \\ 
\leftline{MnCl$_{3}$} &1.62      &3.95&8 &0    \\ 
\leftline{MnBr$_{3}$} &1.76      &4.02&8&0 \\ 
\leftline{MnI$_{3}$}  &1.90       &4.05&8&0\\ 
\hline \\
\leftline{FeF$_{3}$}  &-2.57      &4.15&0&5.10     \\ 
\leftline{FeCl$_{3}$} &-1.68      &4.02&0&3.03   \\ 
\leftline{FeBr$_{3}$} &-2.33     &3.94&0&2.57   \\ 
\leftline{FeI$_{3}$ } &-1.47      &3.82&0&1.83   \\ 
\hline \\
\leftline{NiF$_{3}$}  &76.7       &0.86&2&0      \\                                 
\leftline{NiCl$_{3}$} &43.9       &0.94&2&0     \\ 
\leftline{NiBr$_{3}$} &51.3       &1.02&2&0     \\ 
\leftline{NiI$_{3}$}  &46.7       &1.09&2&0    \\             	 
\hline
\hline
\end{tabular}
\label{Table2} 
\end{table}

\begin{figure*}%[!ht]
\centerline{\includegraphics[width=0.8\linewidth]{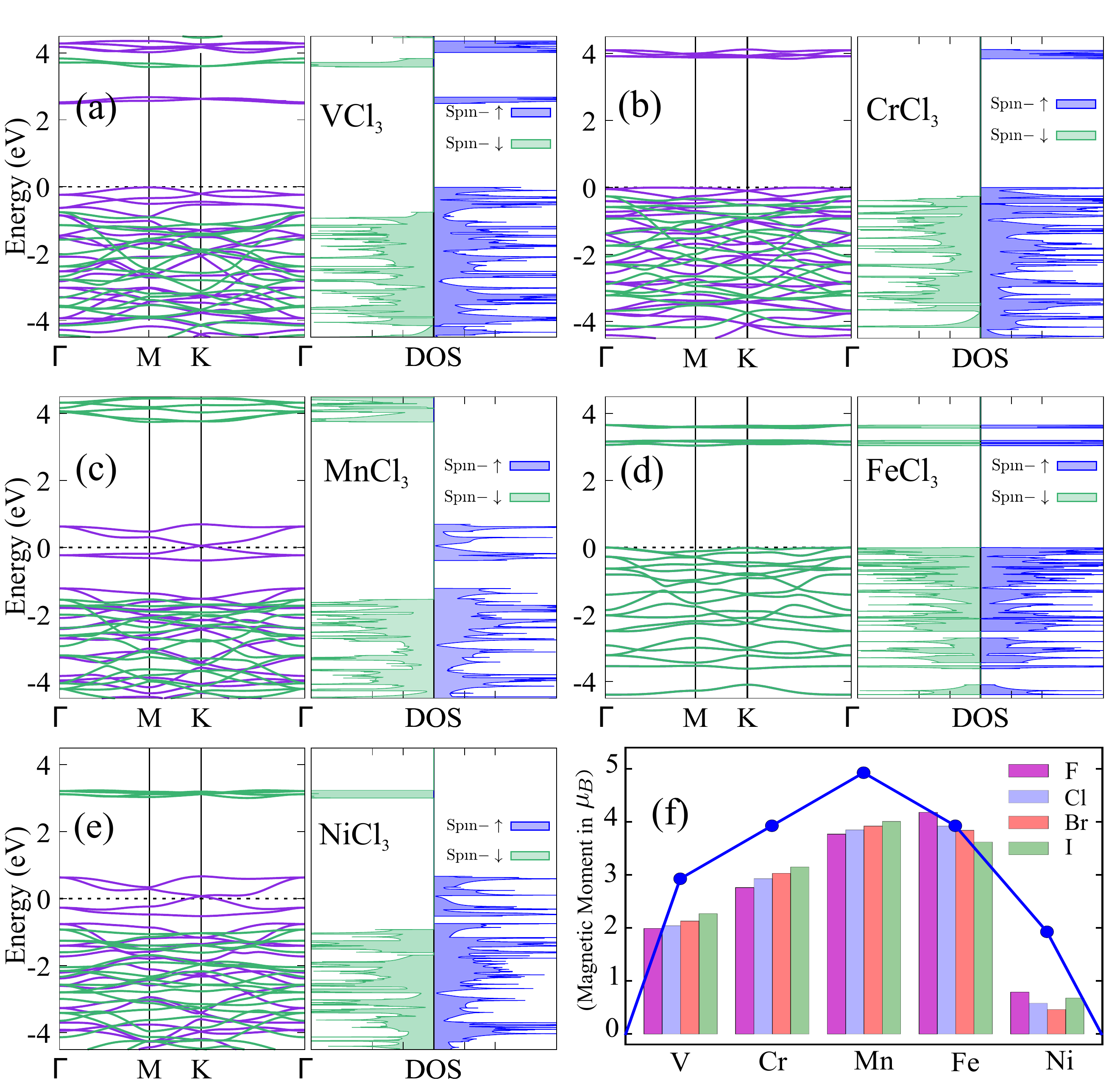}}
\caption{Spin polarized electronic band structure (using HSE06) along the lines joining the high symmetry points in the first Brillouin zone and spin projected density of states (DOS) for monolayer of (a) VCl$_3$, (b) CrCl$_3$, (c) MnCl$_3$, (d) FeCl$_3$, (e) NiCl$_3$ in their magnetic ground state. (f) Magnetic moments of M-atoms in MX$_3$ monolayer. }
\label{fig4}
\end{figure*}

\subsection{Electronic and Magnetic Ground States}
Using the optimized crystal structures (see Table~\ref{Table1} for details), the electronic band structures of MX$_3$ monolayers are calculated and all of the 20 monolayers  are found to have a magnetic ground state. In all MX$_3$ with M$=$ V, Cr, Mn and Ni, the ferromagnetic ground state is obtained, while in case of Fe the anti-ferromagnetic ground state is preferred.
%Comparing the energy difference between the non-magnetic and magnetic state ($E_{NM}-E_{FM/AFM}$), we find that fluorides have the highest difference between the magnetic and non-magnetic ground states in every family (see Fig.~\ref{Fig3}). Other than the fluorides, rest of the trihalides in Mn and Cr family have relatively higher energy difference between the magnetic and non-magnetic ground state, followed by V and Fe family, while Ni family have the lowest value.  Note that, in some of the monolayers, the energy difference between the ferromagnetic and the anti-ferromagnetic ground states is of the order of ~100 meV, although it looks small in Fig.~\ref{Fig3} as the y-axis is in eV scale.

The HSE06 based electronic band structure and the spin projected density of states of one representative tri-halide (chosen to be tri-chloride) in its magnetic ground state is shown in Fig.~\ref{fig4}(a)-(e) and their magnetic moments are compared in Fig.~\ref{fig4}(f). For a complete picture, please refer to 
the Supplementary Information (SI), where a comparison between the PBE and HSE06 based electronic band-structure and spin projected density of states are presented for all the tri-halides studied in this paper. Both the functionals predict qualitatively similar band-structures with bandgap underestimated by PBE, as expected, in case of all the MX$_3$ monolayer families, except for monolayer VX$_3$. This is discussed in detail below. 

\begin{figure}
\centering
\includegraphics[width=\linewidth]{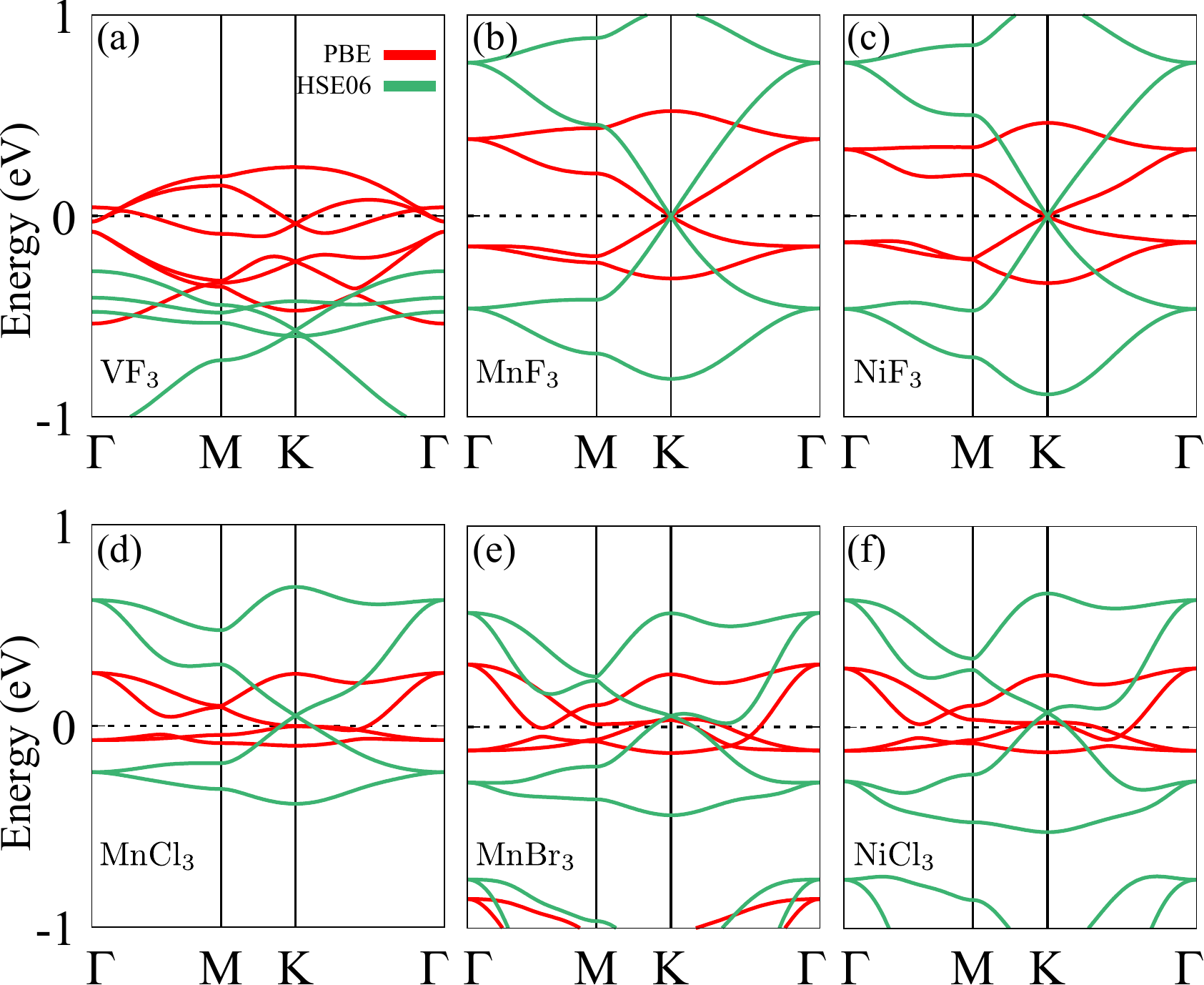}
\caption{The spin polarized Dirac half metallic band structure along the high symmetry directions ($\Gamma$-K-M-$\Gamma$)  for (a) VF$_3$ and (b) MnF$_3$, (c) NiF$_3$, (d) MnCl$_3$, (e) MnBr$_3$ and (f) NiCl$_3$. Only the $\uparrow$-spin bands are shown for clarity. The $\downarrow$-spin bands are gapped, as shown in Fig.~\ref{fig4}, making all of them Dirac half metals.The Fermi level is shown with black dashed line and is set at 0 eV.  
\label{fig5}}
\end{figure}

We find that the electronic and magnetic ground states of all the tri-halides belonging to a particular family of transition metal are similar [see Table~\ref{Table2}]. Our calculations reveal that tri-halides based on Mn and Ni have a ferromagnetic ground state with either metallic or semi-metallic character with a spin polarized Dirac cone. As shown in Fig.~\ref{fig4} (c) and (e), MnCl$_3$ and NiCl$_3$ have very clean spin polarized doubly degenerate Dirac node at the Fermi energy, with two linearly dispersing $\uparrow$-spin bands crossing each other. They also have relatively low density (linearly vanishing) of states at the Fermi level because of the linearly dispersing (Dirac like) bands in vicinity of the $K$-point. Similar behavior is observed in case of several other members of the family, except for tri-iodides, which clearly show a metallic band-structure, with Fermi level ($E_F$) crossing the electronic bands at several points in the Brillouin zone, resulting in a relatively larger density of states at the $E_F$ (see SI). The $\downarrow$-spin bands in case of each of the Mn and Ni tri-halides are similar to that of insulators, with a surprisingly large energy gap of more than 2 eV (see SI). Hence, tri-halides belonging to the Mn and Ni family can be categorized as Dirac half-metals or half-semimetals (based on the character of the $\uparrow$-spin band), as the $\downarrow$-spin states are always gapped.

Similar to Mn and Ni family, the tri-halides based on V are also ferromagnetic. However, they are found to have either a spin polarized Dirac semimetal or metal like state in PBE calculations (similar to Ref.~[\onlinecite{VCl3}]), as opposed to insulating character in HSE06 calculations (band gap ranging from 1.26 eV to 3.23 eV, see Table~\ref{Table2}). The HSE06 based band structure of VCl$_3$ is shown in Fig.~\ref{fig4} (a), which is clearly different from the gap-less energy spectrum obtained from PBE as well as strongly constrained and appropriately normed (SCAN) \cite{PhysRevLett.115.036402} based calculations (see SI). The hybrid functional calculations push all the up-spin bands lower in energy into the valance band, as opposed to PBE or SCAN calculations.  This behavior is observed in case of all the tri-halides belonging to the V family (see SI). Particularly in case of VF$_3$, PBE and SCAN calculations predict a spin-polarized linearly dispersing band (in vicinity of  the $K$-point) while a large bandgap is observed in the HSE06 calculation. Since the HSE06 calculations differ markedly form PBE, SCAN and PBE + U based results,\cite{VCl3} we believe that the HSE06 functional is not well suited for exploring metallic magnetic states in V based compounds. However, it is not very clear why does the difference appear only in case of V, while other transition metals studied here show qualitatively consistent results (see SI for details). 
%Such a fundamental study is outside the scope of this paper.}

The most remarkable feature observed so far is the existence of several Mn, Ni and V (possibly) tri-halide monolayers, with fully spin polarized bands having very clean 2D massless Dirac like dispersion, similar to that of graphene. In hindsight, the existence of 2D Dirac cone in these materials is not  unexpected, as the transition metal atoms form a honeycomb lattice by themselves, similar to the arrangement of carbon atoms in graphene. Some of these Dirac half metals are shown in Fig.~\ref{fig5} (a)-(f). In particular  MnF$_3$, MnCl$_3$, MnBr$_3$, NiF$_3$, and NiCl$_3$ show very clean Dirac half metal state in both PBE as well as HSE06 calculations. These spin polarized Dirac half metals are very interesting from a fundamental physics point of view, as well as for use in spin current generation and other spintronic applications.\cite{PhysRevLett.109.237207,PhysRevB.92.201403}

Unlike half metallic or half semi-metallic tri-halides of Mn and Ni, Cr based tri-halides are found to be insulating (band gap ranging from 1.85 eV to 5.09 eV, see Table~\ref{Table2}), with a ferromagnetic ground state. This is consistent with the experimental observation of ferromagnetic ground state in CrI$_3$. Electronic band structure of CrCl$_3$ is shown in Fig.~\ref{fig4}(b) and qualitatively similar trends are observed for other members of the Cr family (shown explicitly in SI). Note that, magnitude of band gap in case of $\uparrow$ and $\downarrow$ spin state is different for all the Cr tri-halides.

\begin{figure}
\centering
\includegraphics[width=\linewidth]{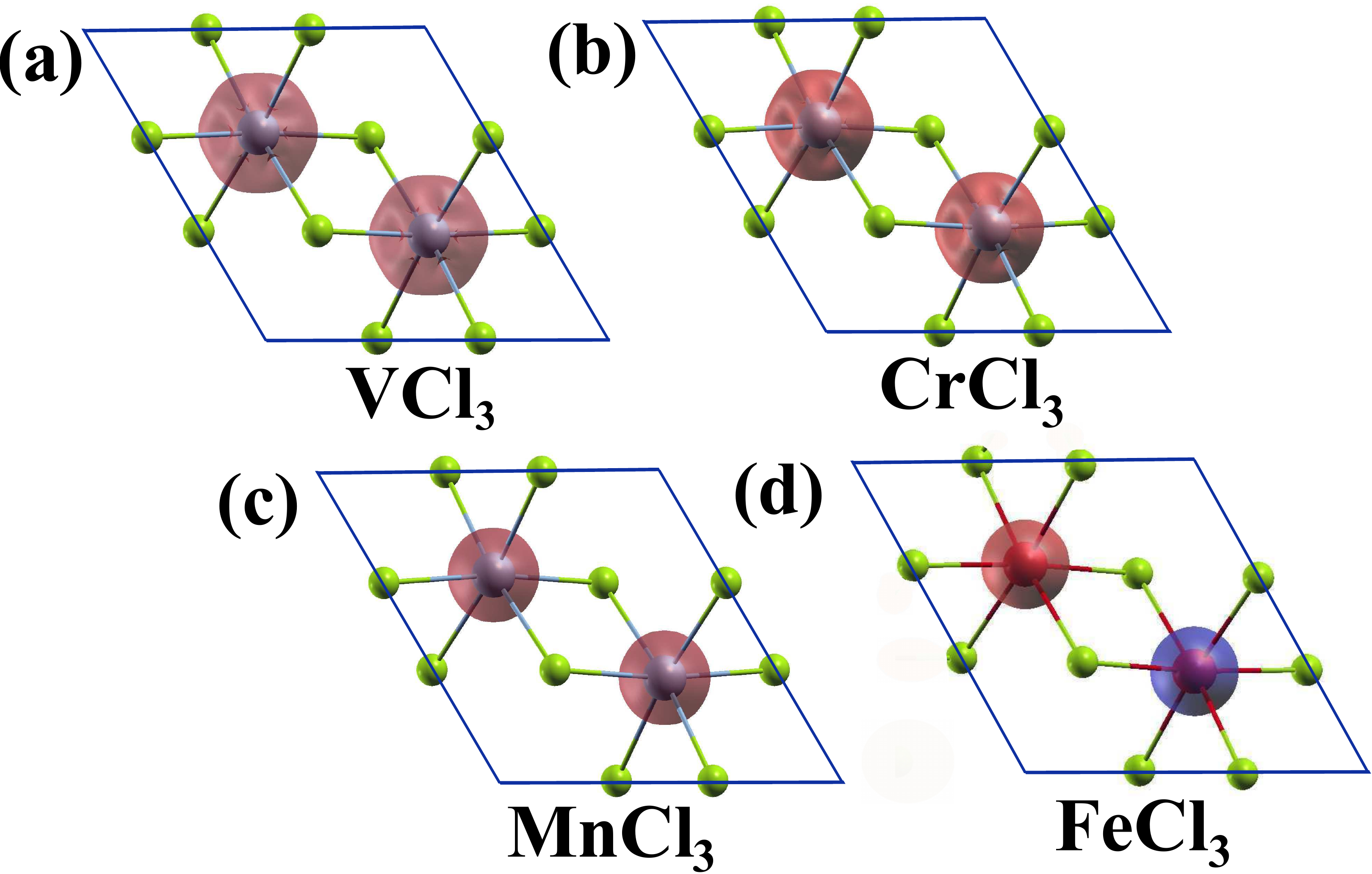}
\caption{Spin-polarized charge densities for (a) VCl$_3$, (b) CrCl$_3$, (c) MnCl$_3$, and (d) FeCl$_3$ monolayer in their ground state. Red denotes the up-spin while blue denotes the down spin in the 3d orbital of the transition metal in MX$_3$ monolayers. Evidently only FeCl$_3$ is anti-ferromagnetic while all the others are ferromagnetic.
% as well as (e) Spin polarized band Structure, total Density of state (DOS) and total Projected Density of state (PDOS) of FeCl$_3$ monolayer in their antiferromagnetic ground state.
\label{fig6}}
\end{figure}

Similar to Cr, Fe-based tri-halides are also insulating (band gap ranging from 1.83 eV to 5.10 eV, see Table~\ref{Table2}). Interestingly, this is the only family with an antiferromagnetic ground state. Electronic band structure of FeCl$_3$ is shown in Fig.~\ref{fig4}(d), which is qualitatively similar for other members of the FeX$_3$ monolayers family as well(see SI). In general, for insulating transition metal tri-halides, magnitude of the bandgap decreases with the size of the halogen atom and nearly 50\% reduction is observed between the two extremes, occupied by the fluoride and iodide, respectively (see Table~\ref{Table2}).

Having discussed the band structure, we now analyze the magnetic properties of the transition metal tri-halides. The calculated values of magnetic moment per transition metal atom for all the MX$_3$ monolayers studied in this paper are illustrated in Fig.~\ref{fig4}(f). Since the transition metal ions are octahedrally coordinated to the halogen atoms, they can exist in either high or low spin state due to crystal field splitting. High crystal field splitting leads to low spin state and vice versa. Based on calculated values of magnetic moment [see Table~\ref{Table2} and Fig.~\ref{fig4}(f)], Mn$^{3+}$ ($\approx 4 \mu_B$) is found to exist in high spin state, while low spin state is observed in case of Ni$^{3+}$ ($\approx 1 \mu_B$). Surprisingly, magnetic moment of iron is found to be $\approx 4 \mu_B$, which is neither high spin ($5 \mu_B$), nor low spin ($1 \mu_B$) state of Fe$^{3+}$. Similar results have been found in a recently published study, which also takes into account the effect of strain, as well as Hubbard-$U$.\cite{Liu_2018} Magnetic moment of vanadium and chromium is found to be consistent with their respective values in the $+3$ oxidation state, which is equal to $2 \mu_B$  and $3 \mu_B$ for V$^{3+}$ and Cr$^{3+}$, respectively. We find that for V, Mn, Cr and Ni based tri-halides the magnetic moment increases marginally with the size of the halogen atom, while in Fe, it decreases slightly from F to I.

Details of the calculated magnetic properties are reported in Table~\ref{Table2}. The ferromagnetic and antiferromagnetic ground states can be identified by non-zero and zero total magnetic moment ($m_T$) per unit cell, respectively. A comparison of absolute magnetic moment ($m_A$) per M atom with $m_T$ [reported in Table~\ref{Table2}, and shown in Fig.~\ref{fig4}(f)] elucidates that the magnetism of MX$_3$ monolayers originates primarily from the transition metal atoms. However, a small ferromagnetic (antiferromagnetic) coupling is observed between the halogen and metal atom in case of fluorides and chlorides (bromides and iodides). The role of the $3d$ transition metal orbitals in the magnetism in MX$_3$ monolayers is further emphasized by plotting the spin densities of V, Cr, Mn and Fe tri-chlorides in Fig.~\ref{fig6} (a)-(d). As discussed previously, only the Fe tri-halides have an anti-ferromagnetic ground state, with magnetic moments of the two Fe atoms in the unit cell pointed in the opposite directions. This is shown for FeCl$_3$ in Fig.~\ref{fig6}(d). 

\begin{figure}%[!ht]
\centering
\includegraphics[width=0.8\linewidth]{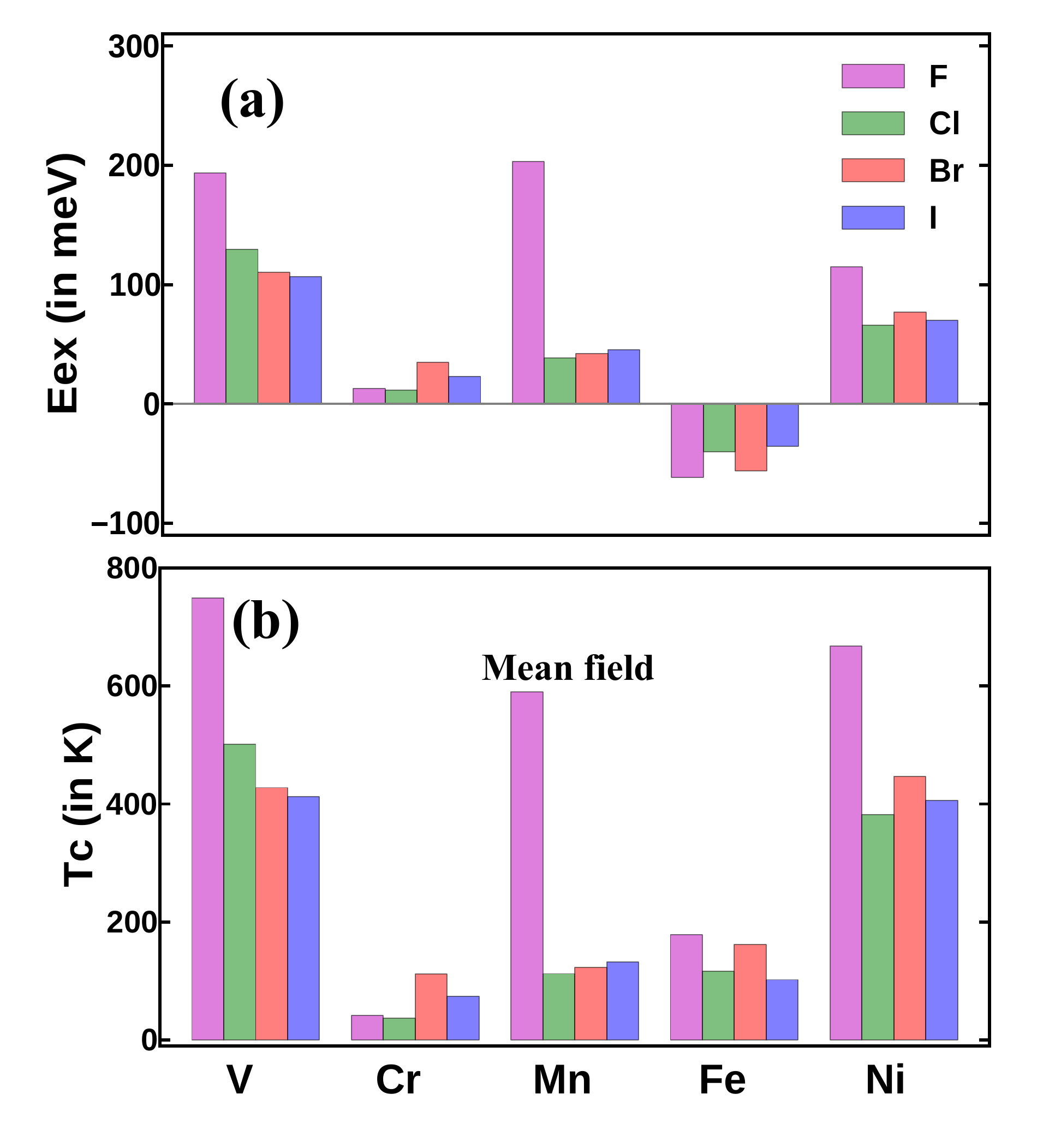}
\caption{(a) A comparison of exchange energies of all the metal tri-halides studied in this paper. While Fe constitutes a family of anti-ferromagnetic tri-halides with negative $E_{\rm ex}$, rest of the transition metal tri-halides are ferromagnetic with positive $E_{\rm ex}$. (b) The corresponding mean field Curie-Weiss transition temperature calculated using the spin-S quantum Ising model.} 
\label{fig7}
\end{figure}

A crude estimate of the exchange coupling and the transition temperature of the magnetic ground state can be obtained by modelling the system as a spin-$S$ quantum Ising model on the honeycomb lattice.  
The spin-$S$ Ising model can be expressed as,
\begin{equation}
\textbf {$H = -\sum_{<ij>}{J_{ij}~ S^z_i}{S^z_j}$}~.
\label{eqham}
\end{equation}  
Here, $J_{ij}$ is the exchange coupling constant between sites $i$ and $j$ sites hosting the magnetic moment of the hybridized transition metal with $S_{i/j}^z  =  m_A/2$. The value of $S^z$ is taken to be equal to 1, 3/2, 2, 2 and 1/2 for M = V, Cr, Mn, Fe, and Ni series, respectively (see Table \ref{Table2} and Fig.~\ref{fig4}(f)). As discussed in the previous paragraph, we find a relatively weak ferromagnetic or anti-ferromagnetic M-X coupling, compared to the much stronger coupling between the two neighboring M atoms. Thus, for simplicity we ignore the M-X coupling, and consider only an isotropic nearest neighbor (in M-M) Ising Hamiltonian in Eq.~\ref{eqham}.

In an isotropic model with nearest neighbor interactions, $J_{ij} = J$ and the sign of $J$ dictates the magnetic ordering in the crystal. Positive value of $J$ leads to ferromagnetic ordering and negative value of $J$ results in anti-ferromagnetic configuration.  Using the isotropic spin-$S$ Ising model with nearest neighbor interactions, $J$ is given by \cite{doi:10.1063/1.2145878,NiCl3}
\begin{equation}
\textbf {$J = \frac{E_{\rm ex}}{2z_0(S^z)^{2}}$}.
\label{eqex}
\end{equation}
Here $z_0 =3$ is the co-ordination number of the M-atoms forming a honeycomb lattice. The exchange energy in Eq.~\ref{eqex} is defined as the half of the of the energy difference between the anti-ferromagnetic and ferromagnetic state.\cite{doi:10.1063/1.2145878}  Calculated values of E$_{\rm ex}$ for all the MX$_3$ monolayers are compared in Fig.~\ref{fig7}(a). We find that fluorides have higher value of exchange energy than rest of the trihalides, Cr-family being the only exception. In general, the V based tri-halide family have reasonably large value of E$_{\rm ex}$, followed by Ni, Mn, Fe and Cr tri-halides.

The magnetic transition temperature ($T_c$) can be estimated using the Curie-Weiss mean field theory for a quantum spin-$S$ Ising model.  For $T > T_c$, the ordered magnetic state is completely destroyed by thermal energy, yielding a paramagnetic (non-magnetic ) state with random spin orientation and zero net magnetic moment in the absence of an external magnetic field. Within the mean field theory, the transition temperature is given by \cite{2015arXiv151103031S}
\begin{equation}
T_c = \frac{z_0 J S^z (S^z+1)}{3 k_B}~,
\end{equation}
where $k_B$ denotes the Boltzmann constant. Calculated values are illustrated in Fig.~\ref{fig7}(b). As shown in the figure, the VX$_3$ and the NiX$_3$ families have relatively higher transition temperatures, ranging from 400 to 800 K.  In case of MnX$_3$ and FeX$_3$ families, $T_c$ lies in the range of 100 to 200 K (other than MnF$_3$, having a $T_c\sim 600$ K). Other than the bromide, members belonging to the CrX$_3$ family have sub 100 K $T_c$ values, which are lower than rest of the trihalide families. Note that, the mean-field theory is known to over-estimate the $T_c$. For example, while experimentally observed $T_c$ value of CrI$_3$ is reported to be 45 K,\cite{Nature_Cri3} our calculations  predict it to be 75 K. However, the relative trend between the $T_c$ of different materials should more or less be along the lines of Fig.~\ref{fig7}(b). 

\section{\label{sec:level6}Conclusion}
Motivated by the recent discovery of two dimensional magnetic materials, we do a comparative study of several transition metal tri-halide  monolayers based on {\it ab-initio} calculations. We find that the tri-halides of V, Mn, and Ni are either metallic or semimetallic with primarily a ferromagnetic ground state. Cr based tri-halides are insulators with a ferromagnetic ground state. Fe based tri-halides are anti-ferromagnetic insulators. We also predict VX$_3$ and NiX$_3$ family of tri-halides to have relatively high magnetic transition temperature. The transition metal tri-halides, MnF$_3$, MnCl$_3$, MnBr$_3$, NiF$_3$, NiCl$_3$ and VF$_3$ are found to host very clean spin polarized Dirac half metallic states. These spin polarized Dirac half metals are very useful for spintronic applications and devices.  Amongst these, VF$_3$, NiCl$_3$ and MnF$_3$ monolayers may even be in thermodynamically stable ferro-magnetic state at room temperature. 

\section{Acknowledgments}
The authors acknowledge funding from the Ramanujan fellowship research grant, the DST Nanomission project and SERB (EMR/2017/004970). The authors also thank the computer center of IIT Kanpur for providing HPC facility.

%%%&&&&&&&&&&&&&&&&&&&&&&&&&&&&&&&&&&&&&&&&&&&&&&&&&&&&&&&&&&&&&&&&&&&&&&&
%\bibliographystyle{IEEEtran}
%\bibliographystyle{IEEEtran}
%
\bibliography{ref}

\end{document}